\documentclass[twocolumn,showpacs]{revtex4}
\usepackage{epsfig}
\usepackage{amssymb}

\begin{document}
\title{Probing Nonequilibrium Fluctuations through Linear Response}
\author{Takahiro Sakaue\footnote{E-mail: sakaue@scphys.kyoto-u.ac.jp} and Takao Ohta}
\affiliation{Department of Physics, Graduate School of Science, Kyoto University, Kyoto 606-8502, Japan}

\begin{abstract}
Linear response analysis in the nonequilibrium steady state (Gaussian regime) provides two independent fluctuation-response relations. One, in the form of the symmetric matrix, manifests the departure from the equilibrium formula through the quantity so-called {\it irreversible circulation}. The other, in the anti-symmetric form, connects the asymmetries in the fluctuation and the response function. 
These formulas represent characteristic features of fluctuations far from equilibrium, which have no counterparts in thermal equilibrium.
\end{abstract}

\maketitle

It is now well recognized that the macroscopic response property of a system near thermal equilibrium is closely related to the dynamical fluctuations of its constituent elements in the scale down to the mesoscopic to molecular size.
This striking connection is formulated in terms of the linear response theory and known as the fluctuation-dissipation theorem (FDT)~\cite{Kubo}.
The theorem is a consequence of the microscopic reversibility and directly linked to the symmetry of the response function in the system with multiple degrees of freedom, i.e., Onsager's reciprocal relation for the transport coefficients~\cite{Onsager}.
These concepts play a crucial role for our understanding the dynamic hierarchical structure of nature.

Away from equilibrium, however, the FDT is generally no longer valid and associated asymmetries show up as a key feature of the nonequilibrium state. This statement is true even for the system characterized by Gaussian fluctuations with the broken time-reversal symmetry, in which the absence of the detailed balance does not allow the characterization of fluctuations from the measured response function.
Indeed, recent several studies have reported the involved fluctuation-response relation in nonequilibrium systems with negligible nonlinearity, in which the equilibrium FDT formula is broken in such a way that the correlation and the response function evolves differently with time. Examples include the driven colloids~\cite{driven_colloid}, dissipative systems, i.e., granular materials~\cite{granular1, granular2}, and some phase-ordering systems under shear flow~\cite{phase_ordering}. Here the origin of the complexity arises not from the nonlinearity but from the coupling between different degrees of freedoms.

The aim of the present paper is to elucidate the fluctuation-response relation in nonequilibrium Gaussian regime by focusing on {\it asymmetries in cross correlations}. We shall show that there is no complication due to the different temporal dependence in a {\it matrix} representation and all the nonequilibrium effects appear as the FDT ratio {\it matrix}, which consists of the intensity of the noise and the so-called {\it irreversible circulation of fluctuation} as a manifestation of the violation of the detailed balance.

To survey the problem under consideration, we first employ a simple polymer model under shear flow, for which both the correlation and the response function can be calculated easily. We then proceed to the general argument based on the linear response analysis applied to the nonequilibrium steady state, in which the dynamics of fluctuations obey Gaussian statistics. The characteristic of nonequilibrium fluctuation-response relation is nicely demonstrated by decomposing it into symmetric and anti-symmetric parts. In particular, the anti-symmetric part of the response concerns the deviation from the reciprocal relation and one can prove the exact relationship between it and the nonequilibrium component of the fluctuation. We then argue that the results persist even to the nonlinear dynamics provided that the fluctuation around the secular motion is Gaussian, as is usually expected for macroscopic systems.

{\it Polymer in shear flow:}
Let us consider two beads connected by a harmonic spring in the thermal bath of the temperature $T$. Although simple, this model, called dumbbell model, enables one to capture basic rheological properties of polymer solutions~\cite{Larson}. Being placed in, for instance, a shear flow, the model provides one of the simplest examples, in which the detailed balance condition is violated. 
The extension to more realistic model with internal modes is straightforward by introducing Rouse modes. Each bead bears a dipole, or simply the electric charge $\pm q$ of opposite signs at both ends so as to be manipulated externally by the time dependent electric field ${\vec E}(t)$. The equation of motion for the end-to-end distance ${\vec x} =(x, \ y, \ z)$ of the dumbbell in flow field is given by
\begin{eqnarray}
\gamma ({\dot x}_i -\kappa_{ij} x_j) = -k x_i + w_i(t) + f^{(p)}_{i}(t)
\label{dumbbell_equation}
\end{eqnarray}
where  $\gamma$, $k$ are the friction and spring constant, which set the relaxation time $\tau_0 = \gamma/k$, ${\vec f}^{(p)}(t)=2q {\vec E}(t)$ is the probing force and ${\vec w}(t)$ is the Gaussian distributed random force with zero mean and the variance $\langle w_i(t)  w_j(t') \rangle = 2 \gamma T \delta_{ij} \delta (t-t')$. 
The bracket denotes the statistical average in the steady state and the repeated indices imply the summation. The Boltzmann constant is set to be unity throughout the paper. We set the direction of the flow and the velocity gradient to be $x$ and $y$ axes, respectively, i.e., the velocity gradient tensor $\kappa_{i j}={\dot \gamma} \delta_{1 2}$. Since the vorticity ($z$) direction is not affected by the flow, we concentrate on the dynamics in $x-y$ plane.
Equation.~(\ref{dumbbell_equation}) is easily solved, leading to the steady state correlation function $ C_{i j}(t-t')= \langle x_{i}(t) x_{j}(t') \rangle$ in the absence of the external field;
\begin{eqnarray}
{\mathbf C}(t) 
= \frac{T}{k}e^{-\frac{t}{\tau_0}}\left( \matrix{
1+\frac{W^2}{2}(1+\frac{t}{\tau_0}) & W(\frac{1}{2}+\frac{t}{\tau_0}) \cr
\frac{W}{2} &1
}
\right) 
\label{dumbbell_C}
\end{eqnarray}
where the dimensionless shear rate $W= {\dot \gamma} \tau_0$ is introduced.
The response property is examined by switching on the constant electric field at time $t=t_0$ and following the average evolution of the dumbbell toward the new stable state;
\begin{eqnarray}
\langle x_i(t-t_0) \rangle_p=R_{ij}(t-t_0) f^{(p)}_{j}
\end{eqnarray}
where the subscript $p$ indicates the presence of the probe force and $R_{ij}(t) \equiv \int_0^t X_{ij}(s) ds$ is the integrated response:
\begin{eqnarray}
{\mathbf R}(t)
=\frac{1}{k} \left( \matrix{
1-e^{-\frac{t}{\tau_0}} & W[1-e^{-\frac{t}{\tau_0}}(1+\frac{1}{\tau_0})] \cr
0 &1-e^{-\frac{t}{\tau_0}} 
}
\right) 
\label{dumbbell_R}
\end{eqnarray}

\begin{figure}
\includegraphics[width=6cm]{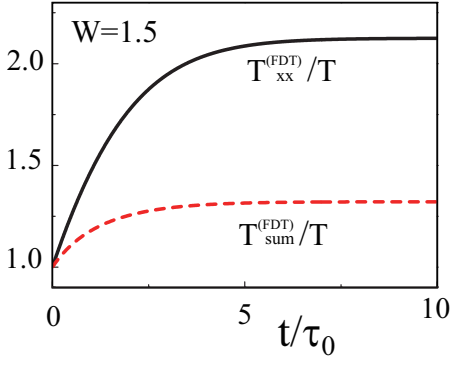}
\caption{Time evolutions of the FDT ratio ($x$ component). The ratio is sometimes defined using the sum of all components; $T^{{\rm (FDT)}}_{sum}(t)=- d (\sum_{\alpha \beta}X_{\alpha \beta})/d(\sum_{\alpha \beta}C_{\alpha \beta})$. Both asymptotically approach the static values $T^{{\rm (FDT)}}_{xx}(t) \rightarrow T(1+W^2/2)$ and $T^{{\rm (FDT)}}_{sum}(t) \rightarrow T(1+W^2/(2(2+W))$.}
\label{plot}
\end{figure}

At equilibrium ($W=0$), these quantities are connected through FDT;
\begin{eqnarray}
TX_{i j}(t)=-\frac{d}{dt}C_{i j}(t)
\label{FDT_equ}
\end{eqnarray}
or equivalently $T R_{ij}(t) = C_{ij}(0)-C_{ij}(t)$. Now, a shear flow breaks the detailed balance and the nonequilibrium effect is often quantified by the FDT ratio $T^{{\rm (FDT)}}_{\alpha \beta}(t)=- dX_{\alpha \beta}/dC_{\alpha \beta}$.
In Fig.~(\ref{plot}), we plot the FDT ratio of the dumbbell under shear. The same, or similar plots were reported in various systems~\cite{driven_colloid, granular1, granular2}.
Individual components of the response and the correlation evolve differently with time, thus, the FDT ratio is time-dependent in short time and approaches to the static value ($>T$) at longer time $t >> \tau_0$. Our crucial observation is that the relation can be cast into the following matrix form:
\begin{eqnarray}
\chi_{i j}(t)  \Theta_{j k} = -\frac{d}{dt}C_{i k}(t)
\label{FRR}
\end{eqnarray}
where the matrix $\Theta_{ij}$ has the dimension of temperature and expressed as
\begin{eqnarray}
{\mathbf \Theta} \equiv [{\mathbf \Sigma}  {\tilde {\mathbf X}}(0)]^{-1}=T\left( \matrix{
1 & -\frac{W}{2} \cr
\frac{W}{2} & 1
}
\right)
\label{dumbbell_theta}
\end{eqnarray}
with $\Sigma_{ij} = [C_{ij}(0)]^{-1}$
 and the static susceptibility ${\tilde X}_{ij}(0)$ (see eq.~(\ref{susceptibility})).
 Although compact, this formula is yet useless because of the presence of the involved quantity $\Theta_{ij}$. This matrix, while reduces to the bath temperature (thus, scalar quantity), i.e., $\Theta_{ij} = T \delta_{ij}$ at equilibrium, is responsible for all the nonequilibrium effect, which makes eq.~(\ref{FRR}) distinct from eq.~(\ref{FDT_equ}).
 We shall explore the meaning of this factor in the general linear stochastic system discussed below.

{\it General model}:
Let us consider a classical system (with $n$ gross variables) coupled to a heat bath at temperature $T$.
In the Markovian level of description, the dynamical variable ${\vec x}= (x_1, x_2, \cdots, x_n)$ obeys the following Langevin equation;
\begin{eqnarray}
{\dot x}_i(t) = -K_{ij}  x_j(t) + \xi_i(t) + v^{(p)}_i(t)
\label{basic_eq}
\end{eqnarray}
where the matrix $K_{ij}$ represents the regression to the secular motion, ${\vec \xi}(t)$ is a random force and a weak perturbation ${\vec v}^{(p)} (t)$ is switched on when probing the response property of the system.
A random force is assumed to be Gaussian white noise with zero mean and the variance $\langle \xi_i(t) \xi_j(t') \rangle = 2 D_{ij} \delta (t-t')$.
The regression matrix can be generally decomposed into two parts $K_{ij} = K^{(1)}_{ij} + K^{(2)}_{ij}$.
The first is associated with the conservative force
$K^{(1)}_{ij}  x_j =  L_{ij} \ \nabla_j V({\vec x})$
where $L_{ij}$ is a mobility matrix and 
$V({\vec x}) = (1/2) U_{ij} x_i x_j$
is a potential energy.
The matrices $L_{ij}$ and $U_{ij}$ are both symmetric.
The second part $K^{(2)}_{ij}$, on the other hand, represents the contribution from non-conservative forces, which drives the system out-of-equilibrium. 
The steady state distribution function is
\begin{eqnarray}
P({\vec x}) = \frac{\exp{[-\phi({\vec x})]}}{\int d{\vec x} \exp{[-\phi({\vec x})]}}
\end{eqnarray}
with a quadratic generalized potential $\phi({\vec x})=(1/2 )\Sigma_{ij} x_i x_j$.

Now we switch on the probe force $f^{(p)}_i (t) = L_{ij}^{-1} v^{(p)}_j(t)$. The linear response of the system is described by
\begin{eqnarray}
\langle x_i(t)\rangle_p  = \int_{-\infty}^{t} dt' X_{ij}(t-t') f^{(p)}_j(t')
\end{eqnarray}
where $X_{ij}(t)$ is a response function.
If one keeps applying a constant force ${\vec f}^{(p)}$ for an enough time, the system will eventually settle in a new steady state with the average and the distribution function given, respectively, by
\begin{eqnarray}
\langle x_i \rangle_p  = {\tilde  X_{ij}}(0)  f^{(p)}_j
\end{eqnarray}
and
\begin{eqnarray}
P_p({\vec x}) = \frac{\exp{[-\phi_p({\vec x})]}}{\int d{\vec x} \exp{[-\phi_p({\vec x})]}}
\label{P_p}
\end{eqnarray}
where the static susceptibility is defined by (footnote~\footnotemark[1]) \footnotetext[1]{Note that the complex admittance is ${\tilde X}_{ik}(\omega) = (K_{ij}-i \omega \delta_{ij})^{-1}L_{jk}$ from eq.~(\ref{basic_eq}).}
\begin{eqnarray}
{\tilde X_{ik}}(0) = \int_0^{\infty} dt \  X_{ik}(t) = K^{-1}_{ij} L_{jk}
\label{susceptibility}
\end{eqnarray}

By expanding $\phi_p=(1/2 )\Sigma_{ij} (x_i- \langle x_i \rangle_p) (x_j- \langle x_j \rangle_p)$ and retaining only a linear term in ${\vec f}_p$, one obtains
\begin{eqnarray}
\phi_p({\vec x}) = \phi({\vec x}) -  {\tilde X}_{ij}(0) \nabla_i \phi({\vec x}) f^{(p)}_j + O(\{{\vec f}^{(p)}\}^2)
\label{phi_expand}
\end{eqnarray}
Putting eq.~(\ref{phi_expand}) into eq.~(\ref{P_p}) gives the following relation between $P({\vec x})$ and $P_p({\vec x})$ in the linear response regime:
\begin{eqnarray}
P_p({\vec x}) &\simeq& P({\vec x}) \left[ 1+ {\tilde X}_{ij}(0) \nabla_i \phi({\vec x}) f^{(p)}_j \right] \nonumber \\
&=& P({\vec x}) \left[ 1+   \Theta_{ij}^{-1}  x_i f^{(p)}_j \right]
\label{P_p_expand}
\end{eqnarray}
with the matrix $\Theta_{ik} = [\Sigma_{ij}  {\tilde X}_{jk}(0)]^{-1}$ introduced in eq.~(\ref{dumbbell_theta}).

To examine the dynamic response, we then turn off the probe force, say, at $t=0$, and follow the relaxation to the original steady state.
During this relaxation process, the average is expressed as
\begin{eqnarray}
<{\vec x}(t)>_{relax} = \int d{\vec x} \ {\vec x} P({\vec x}, t)
\label{x_t_relaxation}
\end{eqnarray}
The probability distribution function at time $t$ is connected with the initial ($t=0$) one through
\begin{eqnarray}
P({\vec x}, t) = \int d{\vec x}' G({\vec x}, {\vec x}'; t) P({\vec x}', 0)
\end{eqnarray}
where $G({\vec x}, {\vec x}'; t)$ is a propagator of the Fokker-Plank equation corresponding to eq.~(\ref{basic_eq}).
By noting $P({\vec x}, 0)=P_p({\vec x})$ and using eq.~(\ref{P_p_expand}), one can transform eq.~(\ref{x_t_relaxation}) as follows
\begin{eqnarray}
<{\vec x}(t)>_{relax}&=&\int d{\vec x} \int d{\vec x}' \ {\vec x} \ G({\vec x}, {\vec x}'; t) \ P_p({\vec x}')
 \nonumber \\
&=& Q_{ij} (t) f^{(p)}_j 
\end{eqnarray}
with the relaxation function $Q_{ik} (t) = \int_t^{\infty} ds \ X_{ik}(s)   =C_{ij}(t)  \Theta_{jk}^{-1}$.
After differentiation with respect to time, one arrives at the fluctuation-response relation: eq.~(\ref{FRR}).

To establish a clear connection between the response and nonequilibrium fluctuations, let us analyze the quantity $\Theta_{ij}$. The essential feature of the nonequilibrium steady state is the presence of non-vanishing probability current. In Gaussian regime, this corresponds to a circulating flux, which can be invoked by representing the steady state covariance in the form:
\begin{eqnarray}
C_{ik}(0) = K_{ij}^{-1} ( D_{jk} + \Omega_{jk})
\label{C0_NESS}
\end{eqnarray}
where the anti-symmetric matrix
\begin{eqnarray}
\Omega_{ik} \equiv \frac{1}{2}[K_{ij}C_{jk}(0) - C_{ji}(0)K_{kj}]
\label{Omega_dif}
\end{eqnarray}
represents the deviation from the Onsager reciprocity for the kinetic coefficient and called irreversible circulation of fluctuation~\cite{Tomita-Tomita}.
The steady state probability current $j_i({\vec x}) = \Omega_{ij} \nabla_j P({\vec x})$ is generally a function of ${\vec x}$ and divergenceless ($\nabla_i  j_i({\vec x})=0$).
Using eqs.~(\ref{dumbbell_theta}),~(\ref{susceptibility}) and ~(\ref{C0_NESS}), the quantity $\Theta_{ij}$ is decomposed into two constituents;
\begin{eqnarray}
\Theta_{ik} = L_{ij}^{-1} (D_{jk}+ \Omega_{jk})
\label{Theta}
\end{eqnarray}
Equation~(\ref{Theta}) clearly shows two requisites for the thermal equilibrium: 1) a property of noise $D_{ij}=T L_{ij}$, i.e., the relation known as FDT of 2nd kind, and 2) no presence of the probability current $\Omega_{ij}=0$.

Now we proceed to express eq.~(\ref{FRR}) in a frequency space.
Let us first Fourier-Laplace transform eq.~(\ref{FRR}):
${\tilde X}_{ij}(\omega)   \Theta_{jk} = C_{ik}(0) + i \omega {\tilde G}_{ik}(\omega)$
,where we have introduced ${\tilde X}_{ij}(\omega) = \int_0^{\infty} dt \ X_{ij}(t) \ e^{i\omega t}$ (complex admittance) and ${\tilde G}_{ij}(\omega) = \int_0^{\infty} dt \ C_{ij}(t) \ e^{i\omega t}$.
Using the property of the correlation function $C_{ij}(t) = C_{ji}(-t)$ (as a consequence of the time translational invariance), the power spectrum is expressed as 
${\tilde C_{ij}}(\omega) = \int_{-\infty}^{\infty} dt \ C_{ij}(t) \ e^{i \omega t}
= {\tilde G}_{ji}^{*}(\omega) + {\tilde G}_{ij}(\omega)$
(the symbol $A_{ij}^{*} $ denotes the complex conjugate of $A_{ij}$.)
Then, it follows that
\begin{eqnarray}
\Delta_{ij}^{(s)}[{\tilde{\mathbf X} }''(\omega) {\mathbf \Theta}]= \omega {\tilde C}_{ij}'(\omega) 
\label{FRR_w1}
\end{eqnarray}
\begin{eqnarray}
\Delta_{ij}^{(as)}[{\tilde{\mathbf X} }'(\omega) {\mathbf \Theta}]= \omega {\tilde C}_{ji}''(\omega) 
\label{FRR_w2}
\end{eqnarray}
where $\Delta_{ij}^{(s)} [{\mathbf A}] = A_{ij} + A_{ji}$ and $\Delta_{ij}^{(as)} [{\mathbf A}] = A_{ij} - A_{ji}$ represent the symmetric and anti-symmetric components of the matrix $A_{ij}$, and prime and double-prime denote the real and imaginary parts, respectively. Note that the imaginary part of the power spectrum is anti-symmetric and connected to the breaking of time-reversal symmetry:
\begin{eqnarray}
{\tilde C}_{ij}''(\omega) ={\mathcal Im}\left[\int_0^{\infty} dt \ [C_{ij}(t)-C_{ij}(-t)]\exp{(i \omega t)} \right]
\label{time_reversal_symmetry}
\end{eqnarray}

From now on, we assume the FDT of 2nd kind, i.e., $D_{ij} = T  L_{ij}$. While this implies that the microscopic degrees of freedom representing the heat bath is equilibrated, the FDT of 1st kind is not yet realized in the presence of the non-conservatice force $K^{(2)}_{ij}$.
From this and eq.~(\ref{Theta}), one obtains
 \begin{eqnarray}
\Theta_{ik} = T \delta_{ik} + L_{ij}^{-1}  \Omega_{jk}
\label{Theta_T_Omega}
\end{eqnarray}
By substituting  (\ref{Theta_T_Omega}) into 
eqs. ~(\ref{FRR_w1}) and~(\ref{FRR_w2}), these are written, respectively, as
\begin{eqnarray}
T\Delta_{ij}^{(s)}[{\tilde{\mathbf X} }''(\omega)] -\omega {\tilde C}'_{ij}(\omega) = -\Delta_{ij}^{(s)}[{\tilde {\mathbf X}}''(\omega) {\mathbf L}^{-1} {\mathbf \Omega}]
\label{FRR_w1_2}
\end{eqnarray}
\begin{eqnarray}
T\Delta_{ij}^{(as)}[{\tilde{\mathbf X} }'(\omega)] +\omega {\tilde C}''_{ij}(\omega) = -\Delta_{ij}^{(as)}[{\tilde {\mathbf X}}'(\omega) {\mathbf L}^{-1} {\mathbf \Omega}]
\label{FRR_w2_2}
\end{eqnarray}
It is evident from these expressions that the irreversible circulation makes the fluctuation-response relation quite different from that near equilibrium. That is,
near equilibrium where $\Omega_{ij}=0$, the right-hand side in eq.~(\ref{FRR_w1_2}) is zero (which is referred to as the FDT of 1st kind), while both terms of the left-hand size in eq.~(\ref{FRR_w2_2}) (${\tilde C}_{ij}''(\omega)$ and the anti-symmetric part of ${\tilde X}_{ij}'(\omega)$) equally vanish, i.e., a manifestation of the time reversal symmetry and the reciprocal relation.

Equation~(\ref{FRR_w2_2}) can be simplified further by taking notice of the structure of the anti-symmetric part of the admittance.
Let us here demonstrate it using the simplest case with two gross variables.
The real and imaginary parts of the admittance are generally connected to each other through the Kramers-Kronig relation.
For example, in the case of a harmonic oscillator (with one degree of freedom) in equilibrium: ${\dot x} = -K_1 x + \xi(t)$, the relation is given by ${\tilde X}' \tau_0 \omega = {\tilde X}''$, with a relaxation time $\tau_0 = K_1^{-1}$.
In the present case, one can, indeed, show the following similar relation for the anti-symmetric part~\cite{Sakaue_Ohta}:
\begin{eqnarray}
\Delta_{1 2}^{(as)} [{\tilde {\mathbf X}}'(\omega) {\mathbf \Theta}  ]= -T \omega \tau_{1 2} \Delta_{1 2}^{(as)}[{\tilde {\mathbf X}}''(\omega)]
\label{chi_relation}
\end{eqnarray}
with $\Theta_{ij}$ given by eq.~(\ref{Theta_T_Omega}) and $\tau_{1 2}= \frac{2}{{\rm Tr}[{\mathbf K}]}$.
Combining eq.~(\ref{chi_relation}) with eq.~(\ref{FRR_w2_2}) leads to the following relation between the anti-symmetric parts of the admittance and
the Fourier transform of the cross correlation:
\begin{eqnarray}
T \tau_{1 2} \Delta_{1 2}^{(as)}[{\tilde {\mathbf X}}''(\omega)] = {\tilde C}''_{1 2}(\omega)
\label{chi-C}
\end{eqnarray}
This formula together with eq.~(\ref{FRR_w1_2}), which is the central result of the present paper, connects the nonequilibrium fluctuation, i.e., the part which is odd under the time reversal (eq.~(\ref{time_reversal_symmetry})), to the asymmetry in the response function.

We emphasize that the formulas ~(\ref{FRR_w1_2}),~(\ref{chi_relation}) and~(\ref{chi-C}) provide us with the procedure to explore the properties of the nonequilibrium fluctuations.
If the admittance ${\tilde X}_{ij}'(\omega)$ (or ${\tilde X}_{ij}''(\omega)$) is experimentally accessible, one can obtain the regression matrix $K_{ij}$ and the mobility matrix $L_{ij}$ from its limiting behaviours (see footnote~\footnotemark[1]): ${\tilde X}_{ik} (\omega \rightarrow 0) = K_{ij}^{-1} L_{jk}$ and  ${\tilde  X}_{ij} (\omega \rightarrow \infty) = i L_{ij} \omega^{-1}$.
Then, from eq.~(\ref{chi_relation}) and~(\ref{chi-C}) one obtains the irreversible circulation $\Omega_{ij}$ and the imaginary part of the power spectrum ${\tilde C}_{ij}''(\omega)$, respectively, which are the mostly relevant quantities to characterize the systems far from equilibrium.
Finally, eq.~(\ref{FRR_w1_2}) together with the information so far obtained provides the real part of the power spectrum ${\tilde C}_{ij}'(\omega)$, thus, the complete knowledge of the correlation function.
Although our argument is valid only for Gaussian regime, this does not necessarily exclude the applicability to the nonlinear dynamics. Even when the gross variables obey the nonlinear equation of motion, fluctuations around the secular motion might be still Gaussian, which is usually expected for macroscopic system. In such cases, all the derived results can be applied without any modifications~\cite{Sakaue_Ohta}.

In summary, we have constructed the linear response formalism around the nonequilibrium steady state in Gaussian regime.
Main results are (I) the simple equation,
 (\ref{chi-C}), 
 connecting the nonequilibrium fluctuation to the departure from the reciprocal relation and (II) the equation, 
 (\ref{FRR_w1_2}), 
 quantifying the FDT violation through the irreversible circulation of fluctuation.
These can be shown to be consistent with other works on the fluctuation-response in nonequilibrium steady state, in which the effect of nonlinearity is highlighted\cite{Kurchan2, Harada, Speck_Seifert}.
While restricted, the present discussion in Gaussian regime enable us to establish the useful formulas elucidating connections between essential quantities characterizing nonequilibrium steady state, through which one can extract the role of the coupling between different degrees of freedom away from equilibrium.

We thank T. Ohkuma for useful discussions. The present research is supported by JSPS Research Fellowships for Young Scientists (No. 01263) and by the Grant-in-Aid for the 21st Century COE "Center for Diversity and Universality in Physics" 
and the Grant-in-Aid for the superior area "Soft Matter Physics" both 
from the Ministry of Education, Culture, Sports, Science and Technology (MEXT) of Japan.

\end{document}